\documentclass[useAMS,usenatbib]{mn2e}
\usepackage{epsfig}
\usepackage{graphicx}
\usepackage{psfrag}

\newcommand{\mnras}{Monthly Notices of the Royal Astronomical Society}
\newcommand{\apj}{Astrophysical Journal}
\newcommand{\apjl}{Astrophysical Journal Letters}
\newcommand{\aap}{Astronomy and Astrophysics}

\title[Gravitational Flexion by Elliptical Dark Matter Haloes]{Gravitational Flexion by Elliptical Dark Matter Haloes}
\author[A. J. Hawken and S. L. Bridle]{A. J. Hawken$^{1}$ and S. L. Bridle$^{1}$\thanks{E-mail:
ahawken@star.ucl.ac.uk; sarah.bridle@ucl.ac.uk}\\
$^{1}$University College London, Gower Street, London}

\begin{document}

\date{18/01/09}
\pagerange{\pageref{firstpage}--\pageref{lastpage}} \pubyear{2009}
\maketitle

\label{firstpage}

\begin{abstract}
We present equations for the gravitational lensing flexion expected for an elliptical lens mass distribution. These can be reduced to one-dimensional finite integrals, thus saving significant computing time over a full two-dimensional calculation.
We estimate constraints on galaxy halo ellipticities for a range of potential future surveys, finding that the constraints from the two different types of flexion are comparable and are up to two orders of magnitude tighter than those from shear.
Flexion therefore appears to be a very promising potential tool for constraining the shapes of galaxy haloes from future surveys.
\end{abstract}


\section{Introduction}

Light from distant galaxies is perturbed by the curvature in space-time induced by intervening matter. This distorts the galaxy images and is referred to as gravitational lensing. In the most extreme cases of strong lensing the galaxy images are bent into thin arcs around a concentration of mass, and there may be multiple images of the object. Much more often a galaxy image is distorted only very slightly, and the image is sheared by a simple matrix distortion. Thus an intrinsically circular object will appear as an ellipse. This first-order effect is referred to as weak gravitational lensing.
The next order gravitational lensing effect,
known as ``flexion'', is a relatively new way of probing gravitational lenses
\citep[][]{Goldberg:2001xi, Irwin:2003qm,Bacon, SchneiderEr,2008arXiv0807.1931B}.
Other second-order effects are not expected to be produced by simple gravitational lensing but could be a signature of systematic effects~\citep{TwistTurn}.

Flexion comprises of the one-flexion, or ``displacement'', and three-flexion,
 or ``cardioid shift''. When combined with the shear they describe how ``banana-like" a galaxy behind a gravitational lens appears.
A disadvantage of the conventional weak lensing shear technique is that galaxies frequently have an intrinsically elliptical or stretched shape, and the extra stretching due to lensing is small by comparison. This effectively introduces extra noise into the measurements and requires an average over many galaxies to observe a significant lensing effect. However, galaxies are much less often intrinsically banana-shaped, and thus flexion measurements are expected to have a much higher signal to noise ratio in some regimes. In addition the information obtained is orthogonal to that obtained from shear. Therefore information about lens systems derived from flexion can be combined with that from shear to obtain tighter constraints.

Techniques have been developed to measure flexion from astronomical images,
in the presence of pixelisation and noise~\citep{OUF97,OUF08,GoldbergLeonard,Masseyetal,Irwin:2005nc}.
Flexion has been used to constrain the distribution of mass in clusters of galaxies~\citep{Irwin:2003qm,Leonard, OUF08,Leonard:2008jw} and give indications of the distribution of mass in blank fields~\citep{Irwin:2005nc, Irwin:2006tn}.
Flexion due to the mass distribution in foreground galaxies (galaxy-galaxy flexion) has been observed
in the Deep Lens Survey \citep{GoldbergBacon}. In this paper we discuss the use of galaxy-galaxy flexion to measure halo ellipticities. This builds on the work of~\citet{Bacon}.

A method for using weak gravitational lensing shear to determine galaxy halo ellipticities by stacking many galaxies together was proposed in~\citet{Natarajan:2000im}. This has been used to determine
the ellipticities of galaxy haloes~\citep{Hoekstra:2003pn,Mandelbaum,Parker} and cluster dark matter haloes~\citep{EvansBridle}.
Ellipticities of individual clusters have also been measured~\citep{Cypriano:2003ij,Corless:2008wk}.

Accurate determination of the shapes and orientations of galaxy dark matter haloes can provide constraints on models of galaxy formation and the nature of dark matter.
N-body simulations of non-interacting cold dark matter predict that the these haloes should be triaxial prolate ellipsoids~\citep[e.g.][]{Allgood2006}. Simulations that use cooling gas and dark matter predict that dark matter haloes should be more spherical~\citep[e.g.][]{Kazantzidis2004}. Predictions in the Modified Newtonian Dynamics (MOND) paradigm predict that dark matter haloes should look isotropic at large radii~\citep{MortlockTurner}. Previous weak lensing studies have found flattened, elliptical galaxy dark matter haloes which disfavor the MOND paradigm~\citep{Hoekstra:2003pn,Mandelbaum,Parker}. Thus through measuring the shapes of dark matter haloes, galaxy-galaxy lensing can provide constraints on galaxy formation models and the nature of dark matter.

In Section~\ref{flex} we derive expressions for the components of the flexion produced by an elliptical gravitational lens of arbitrary dimensionless surface density
profile and illustrate the flexion induced by an elliptical Navarro-Frenk-White
\citep[NFW;][]{NFW96,NFW97}) lens. In Section~\ref{surveys} we show predictions for the uncertainties expected on the mean ellipticity of galaxy dark matter haloes. We conclude in Section~\ref{conclusion}.

We adopt a fiducial flat cosmological constant dominated cold dark matter
($\Lambda$CDM) cosmology with $\Omega_{M} = 0.3$, $\Omega_{\Lambda} = 0.7$ and
a fluctuation amplitude parameterized by the root mean square density perturbation in 8 $h^{-1}$ Mpc spheres at the present day in the linear regime, $\sigma_8=0.74 $.

\section{FLEXION BY ELLIPTICAL haloes} \label{flex}

\begin{figure*}
\psfrag{mathcalFT}{\Large{$\mathcal{F}_R$}}
\psfrag{mathcalFX}{\Large{$\mathcal{F}_B$}}
\psfrag{mathcalGT}{\Large{$\mathcal{G}_R$}}
\psfrag{mathcalGX}{\Large{$\mathcal{G}_B$}}
\psfrag{gammaT}{\Large{$\gamma_T$}}
\psfrag{gammaX}{\Large{$\gamma_X$}}
\psfrag{xaxis}{x/arcsec}
\psfrag{yaxis}{y/arcsec}
\includegraphics[width=18cm]{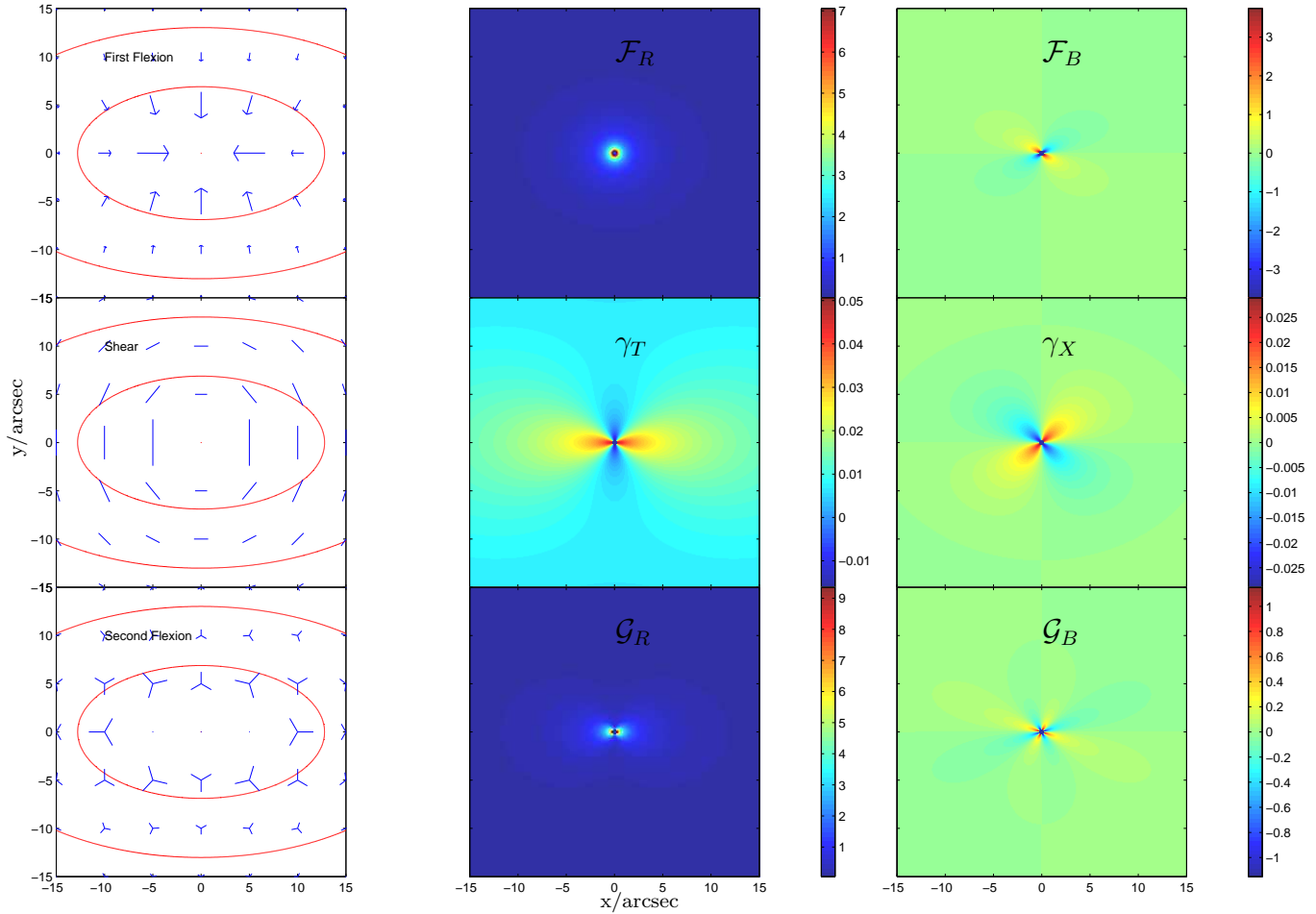}
\caption{Illustrations of the three simplest lensing distortions for an NFW dark matter halo with an ellipticity of $e = 0.3$ and mass of $10^{12}\, h^{-1} M_{\odot}$ and concentration parameter $c = 7.2$. The top left panel shows arrows pointing in the direction of the one-flexion vector, the length
of the arrow is proportional to the magnitude of the one-flexion. The top middle panel is a colour map showing how
$\mathcal{F}_R$ varies across the image the top right panel shows how $\mathcal{F}_B$ varies. The three panels in the middle row illustrate the equivalent quantities for the shear and the bottom row for three-flexion. Colour bars on the right of each flexion image give the distortion per arcsecond. The contours on the left hand panels are at convergence, $\kappa$ values of 0.01 and 0.02. Arrows and sticks that would be distractingly large have been removed.
}
\label{TandX}
\end{figure*}

N-body simulations predict
dark matter haloes to be triaxial ellipsoids
as opposed to spherical. A homeoidal ellipsoid is one in which shells are concentric triaxial ellipsoids all with the same ellipticity and orientation~\citep[e.g.][]{Schramm}.
Thus homoeoidal ellipsoids maintain their symmetry when projected onto a plane.
In practice dark matter haloes are more complicated and may contain substructures and twisted isodensity contours and hence would not be correctly described by a homoeoidal density profile.
However, in galaxy-galaxy lensing many lens haloes are compiled on top of one another, ideally with their major axes aligned. This would smooth out any twisting or substructure, thus a homoeoidal description of the collated density profile is justified.
For a detailed discussion on various ellipsoidal models  and related theorems see \cite{Chandra}.

We now consider flexion by an elliptical mass distribution. An elliptical potential corresponds to an approximately elliptical mass distribution in the small ellipticity limit. For large ellipticities the mass distribution becomes dumbell shaped. \cite{Baltz} get around this by finding a particular sum of elliptical potentials that roughly mimics an elliptical mass distribution. Here we derive the exact equations for an elliptical mass distribution for the first time.

\subsection{First and Second Flexions}

The weak lensing shear and flexion are conveniently described using a complex space formalism introduced to flexion
 by \cite{Bacon} that allows the magnitude and direction of the distortion effects to be expressed simultaneously.
The shear is composed of real and imaginary parts $\gamma = \gamma_1 + i\gamma_2$
where $\gamma_1$ describes the stretching along the $x$ and $y$ axes, and $\gamma_2$ describes stretching along the diagonals of the coordinate system. Similarly we can write for the one-flexion
\begin{eqnarray}
\mathcal{F} &=& \mathcal{F}_1 + i \mathcal{F}_2 = \partial^* \gamma
\end{eqnarray}
and for the three-flexion
\begin{eqnarray}
\mathcal{G} &=& \mathcal{G}_1 + i \mathcal{G}_2 = \partial \gamma .
\end{eqnarray}
where $\partial$ denotes a complex derivative operator $\partial = \partial/\partial x +\ i \partial/\partial y$ with $\partial^*$ as its complex conjugate \citep{Bacon}.

The shear is a second derivative of the lensing potential $\psi$;
$\gamma_1 =\frac{1}{2}(\psi_{xx}+\psi_{yy}) $ and $\gamma_2 = \psi_{xy}$,
where subscripts denote partial differentiation. The real and imaginary components of the flexion can
therefore be expressed explicitly in terms of the third derivatives of the potential
\begin{eqnarray}
\mathcal{F}_1 &=& \frac{1}{2} (\psi_{xxx} + \psi_{yyx} ), \label{F_1}\\
\mathcal{F}_2 &=& \frac{1}{2} (\psi_{xxy} + \psi_{yyy} ), \\
\mathcal{G}_1 &=& \frac{1}{2} (\psi_{xxx} - 3\psi_{xyy} ), \\
\mathcal{G}_2 &=& \frac{1}{2} (3\psi_{xxy} - \psi_{yyy} ). \label{G_2}
\end{eqnarray}
So in order to calculate the flexion components induced at position $(x, y)$ on the image plane by an elliptical halo all one has to do is find derivatives of the potential around an elliptical lens.

By differentiating equations 9-11 of \cite{Keeton} we obtain expressions for the third derivatives of the deflection potential (equations \ref{psi_xxx} - \ref{psi_yyx}) induced by an elliptical dark matter halo of arbitrary lensing convergence, $\kappa$.
In general relativity and the thin-lens approximation, the lensing convergence is simply the mass density projected along the line of sight.
Here we are mapping a one-dimensional $\kappa$ profile, onto homoeoids
with minor to major axis ratio $q$.
We find
\begin{eqnarray}
\psi_{xxx} &=& 6qxK_0 + 4qx^3 L_0 , \label{psi_xxx} \\
\psi_{yyy} &=& 6qyK_2 + 4qy^3 L_3 ,\\
\psi_{xxy} &=& 2qyK_1 + 4qx^2 yL_1 ,\\
\psi_{yyx} &=& 2qxK_1 + 4qy^2 xL_2   \label{psi_yyx}
\end{eqnarray}
Here
\begin{equation}
K_n (x, y) = \int_0^1
\frac{u\,\kappa'\,(\xi (u) )^2\,du} {[1 - (1 - q^2 )u]^{n+1/2}} 
\end{equation}
and
\begin{equation}
L_n (x, y) = \int_0^1
\frac{u^2\,\kappa''\,(\xi(u))^2 du}{[1-(1-q^2)u]^{n+1/2}} 
\end{equation}
are one-dimensional integrals from the centre to
the ellipse
intersecting the $(x,y)$ position of interest. These are given
in terms of the first
and second derivatives of the convergence as a function of ellipse coordinate
\begin{equation}
\kappa'(\xi^2)= \frac{\partial \kappa(\xi^2)}{\partial \xi^2}
\end{equation}
where
\begin{equation}
(\xi (u))^2 = u\bigg(x^2 + \frac{y^2}{[1 - (1 - q^2 )u]}\bigg) 
\end{equation}
is the elliptical coordinate.

By substituting equations \ref{psi_xxx} - \ref{psi_yyx} into \ref{F_1} to \ref{G_2} we can obtain the first and second flexions at any point on the plane $(x,y)$ around an elliptical lens for any $\kappa$. Both \cite{Schramm} and \cite{Keeton} note that only very simple matter distributions give simple analytical relations for elliptical lenses. However, for realistic matter distributions, such as the NFW profile (\cite{NFW96} \cite{NFW97}); the expected flexions around a given lens can easily be computed.

\subsection{ Radial Flexion}

The shear around an elliptical lens can be described in terms of the
tangential $\gamma_T$
and cross $\gamma_X$
components
\begin{eqnarray}
\gamma_T &=& |\gamma | \cos (2\theta_{\gamma}'),\\
\gamma_X &=& |\gamma | \sin (2\theta_{\gamma}')
\end{eqnarray}
where
\begin{equation}
\theta_{\gamma}'= \theta_{\gamma} - \phi + \frac{\pi}{2}
\end{equation}
and $\gamma_1=|\gamma| \cos(2\theta_{\gamma})$, $\gamma_2=|\gamma| \sin(2\theta_{\gamma})$ and $x=r \cos(\phi)$, $y=r\sin(\phi)$ where $r^2 =x^2 +\ y^2$.
Therefore $\theta_{\gamma}$\ is the angle of the shear stretch to the coordinate axes $x,y$, $\theta_{\gamma}'$ is the angle of the shear stretch relative to a rotated set of axes aligned along the radial direction and $\phi$ is the angle subtended at the center of symmetry from the positive x axis to the vector $x,y$ to the position of the background galaxy.

We propose that the equivalent quantities for the flexions should be defined \begin{eqnarray}
\mathcal{F}_R &=& |\mathcal{F} | \cos(\theta_{\mathcal{F}}' ),\\
\mathcal{F}_B &=& |\mathcal{F} | \sin(\theta_{\mathcal{F}}' ), \\
\mathcal{G}_R &=& |\mathcal{G} | \cos(3\theta_{\mathcal{G}}' ), \\
\mathcal{G}_B &=& |\mathcal{G} | \sin(3\theta_{\mathcal{G}}' ),
\end{eqnarray}
where
\begin{eqnarray}
\theta_{\mathcal{F}}'&=&\theta_{\mathcal{F}}-\phi+\pi,\\
\theta_{\mathcal{G}}'&=&\theta_{\mathcal{G}}-\phi,
\end{eqnarray}
and $\mathcal{F}_1 = |\mathcal{F}| \cos(\theta_{\mathcal{F}})$,
$\mathcal{F}_2 = |\mathcal{F}| \sin(\theta_{\mathcal{F}})$,
$\mathcal{G}_1 = |\mathcal{G}| \cos(3\theta_{\mathcal{G}})$ and
$\mathcal{G}_2 = |\mathcal{G}| \sin(3\theta_{\mathcal{G}})$.
Here $\theta_{\mathcal{F}}'$ and $\theta_{\mathcal{G}}'$ are defined in such a way as to make $\mathcal{F}_R$ and $\mathcal{G}_R$ positive
around a circular lens.
These `radial' components are the gravitational contributions of the flexion around a circular lens, any $\mathcal{F}_B$ or $\mathcal{G}_B$ around a circular lens is hence bogus.

We illustrate $ \mathcal{F}_R$, $\mathcal{F}_B$, $\gamma_T$, $\gamma_X$, $\mathcal{G}_R$ and $\mathcal{G}_B$
for an elliptical NFW\ profile in Figure \ref{TandX}. For an elliptical singular isothermal sphere profile the shear is entirely tangential~\citep{1994A&A...284..285K}, just as for a circular lens. We see from the central panel that the result for an NFW\ profile has a small but non-zero cross shear. The cross shear is such that the direction of the shear sticks at positions on the $y=x$ line are slightly more horizontal than a purely tangential line.
They therefore tend towards alignment along the isodensity contours.

The one-flexion has a
significant cross component. This is clearly seen in the top left stick plot, in which the arrows
have a vertical component
and therefore are more closely perpendicular to the isodensity contours.
The three flexion has a small cross component and the far leg of the triangle is close to radial despite the ellitpicity of the lens. The exact perturbation from radial varies according to the angular position on the lens, but is mostly pointing even further from the radial line than the perpendicular to the isodensity contours.

\section{APPLICATION TO SURVEYS} \label{surveys}

\begin{figure}\label{nzEUCLID}
\psfrag{zl}{\large{$z_l$}}
\psfrag{zs}{\large{$z_s$}}
\psfrag{redshift}{Redshift}
\includegraphics[width=8cm]{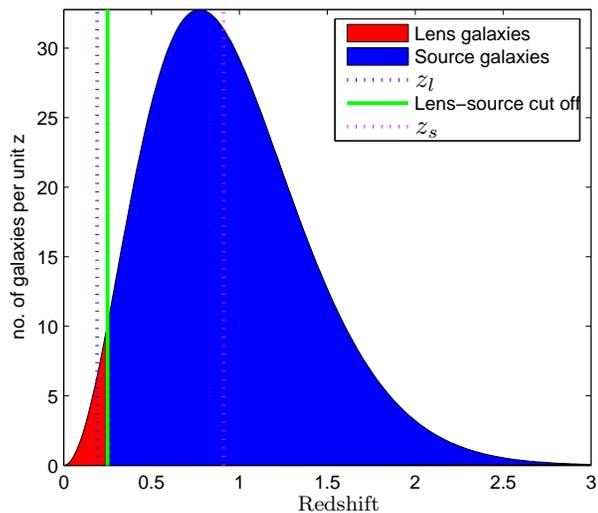}
\caption{Number of galaxies per unit redshift as a function of redshift for a fiducial space-based survey. The green line indicates the cut-off between the lens and source population, shaded red and blue respectively. The blue dotted line shows the median redshift of the lens population whilst the magenta dotted line shows the median redshift of the source population. }
\end{figure}

\begin{figure*}\label{likeEUCLID}
\psfrag{Ellipticity}{Ellipticity}
\psfrag{mathcalF}{\Large{$\mathcal{F}$}}
\psfrag{mathcalG}{\Large{$\mathcal{G}$}}
\psfrag{mathcalLnorm}{\Large{$\mathcal{L}_{\rm norm}$}}
\psfrag{gamma}{\Large{$\gamma$}}
\includegraphics[width=18cm]{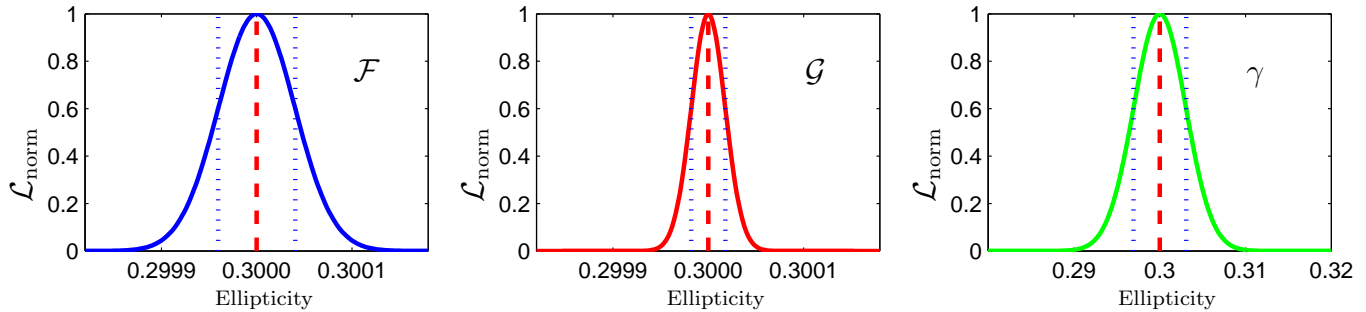}
\caption{Probability as a function of lens ellipticity for the fiducial large area space-based survey. The true ellipticity is shown by the red dashed line and the blue dotted lines show the 68\% confidence limits. The left panel shows the constraints from one-flexion, the central panel for three-flexion and the right hand panel for shear.}
\end{figure*}

Here we estimate the potential for future surveys to measure the ellipticity of haloes using stacked shear and flexion maps. We calculate flexions and shears on a grid
covering one arcminute square centred on the lens centre. This is analagous to stacking galaxies in postage stamps of one arcminute square, centred on the halo center with the major axes aligned along the $x$ axis and averaging observed flexions, or shears, in pixels. This method is most applicable to large scale weak lensing surveys.

\subsection{Lens and source populations}\label{populations}

We need an estimate of the lens parameters, and of lens and source galaxy redshifts, in order to calculate the flexions we would expect to observe at positions around the lens. In this work we carry out an illustrative calculation using mean quantities for each of the lens and source populations.

To estimate the typical redshift of the lens and source we divide up the redshift distribution of galaxies
from Equation 6 of  \cite{BlakeBridle}
\begin{equation}
\frac{dN}{dz} = \alpha \Sigma_0\frac{z^2_l}{z^3_0}  \exp\left(-\left(\frac{z_l}{z_0}\right)^\beta\right)
\end{equation}
where the characteristic redshift
\begin{equation}
z_0 = 0.055(r_{lim} - 24) + 0.39 \\
\end{equation}
and overall surface density
\begin{equation}
\Sigma_0 = \frac{35400}{60^2}\big( \frac{r_{lim}}{24}\big),
\end{equation}
are found from linear fits to their Table 1. Note that this table is derived from observed luminosity functions and extends only up to a magnitude of $24$, therefore the above equations are used to extrapolate beyond the numbers provided. For general ground-based surveys we use $\alpha = \beta = \frac{3}{2}$.

We define the lens population such that there is one lens per square arcminute postage stamp.
The postage stamp size would ideally be small to avoid overlaps between lenses, but to get a good signal from a finite survey area we make a compromise at 1 arcminute. Beyond this the flexion signal is
very noisy. 
We assume that the lens population of interest will be all the low-redshift galaxies, up to some redshift cut-off. We take the cut redshift to be such that there is one lens galaxy per square arcminute.
This means that in our estimation the number of lenses in any survey is solely dependent on the area of that survey whereas the number of sources is dependent upon both the depth and area of the survey. The redshift of the background galaxies is taken to be the median redshift of the population posterior to this cut-off. Similarly the lens redshift is taken to be the median redshift of the lens population.

Using the median redshift of the lens population we can estimate the lower mass limit of the observable haloes at this redshift using the Sheth-Tormen mass function \citep{ShethTormen}:
\begin{equation}
n_h (M )dM = A\bigg(1 + \frac{1}{\nu^ {2q}}\bigg)\sqrt{\frac{2}{\pi}}\frac{\rho_m}{M}\frac{d\nu}{dM}\exp\bigg(-\frac{\nu^2}{2}\bigg)dM
\end{equation}
where $\nu = \sqrt{a} \delta r D(z)\sigma(M )$, $A = 0.322$, $a = 0.707$ and $q = 0.3$.
We do this by varying the lower mass limit until the number density of haloes matches the number density of galaxies for our given magnitude limit. This mass is then used as our lens mas, $M_{200}$, in our calculations of the lensing potential.
Technically this is the lower limit to the mass rather than the mean mass. However it will be close to the mean due to the shape of the mass function. The small difference makes our uncertainty estimates on the conservative side.

Hypothetically any $\kappa$ profile could be plugged into \ref{psi_xxx} - \ref{psi_yyx} but we have chosen an NFW profile because it is
reasonably realistic. The value of the NFW concentration parameter is estimated by assuming that its relationship with $M_{200}$ has the functional form of equation 12 of ~\citet{Seljak2000}
\begin{equation}
c= c_0 \bigg(\frac{M_{200}}{M_*}\bigg)^B,
\end{equation}
where $c_0 = 10$, $B = -0.15$
and $M_*$ is a function of $z$.

As magnitude limit is increased, more low luminosity, low mass, low redshift galaxies become visible. If these are included in the lens population then the median redshift and thus assumed mass of the lens galaxy population will decrease. This decreases the lensing efficiency resulting in little improvement on potential dark matter halo ellipticity constraints as the depth of the survey is increased.
Therefore by default we carry out calculations in which the lens population only contains galaxies brighter than a magnitude limit of 24, even though the survey itself may be deeper. This means that the properties of the lens population are unchanged as the survey depth is increased. This includes the lens cut-off redshift. The source population is then taken to be all the galaxies at redshifts above the lens cut-off redshift and down to the magnitude limit of the survey. The source population redshift then increases with magnitude limit giving a stronger flexion signal.

The lens population from an $r_{lim}=24$ survey covering 20,000 square degrees of sky has a
median redshift of $z_l= 0.19$ and there are $n_s=6\times10^8$ sources with a median redshift of $z_s=0.58$. We find $M_{200}=2.6\times10^{10}M_\odot h^{-1}$ with concentration parameter $c=22$.
An $r_{lim}=26$ survey has $n_s = 3.1\times10^9$  sources with $z_s=0.72$.
Whereas if we were to allow the lens population to change as the survey got deeper, we would have in the $r_{lim}=26$ survey a lens redshift of $z_l= 0.14$, a lens mass of $M_{200} = 9.3\times10^9M_\odot h^{-1}$ with a concentration parameter of $c=26$. There would be $n_s = 3.2\times10^9$ sources with a median redshift of $z_s=0.71$. These subtle changes add up to a noticeable change in ellipticity constraints (Figure ~\ref{maglim}).

We also show results for a fiducial large-area space-based survey covering 20,000 square degrees of sky
with 35 galaxies per square arcminute.
We assume $\alpha=2$, $\beta=\frac{3}{2}$,
$z_0 = 0.63$ and $\Sigma_0 = 27$.
This has lenses with a median redshift of $z_l=0.19$ and $n_s=2.5\times10^9$ sources with a median redshift of $z_s=0.91$. This gives $M_{200}=2.6\times10^10M_\odot h^{-1}$ with $c=22$.

\subsection{Constraints on Lens Ellipticity}\label{constraints on lens ellipticity}

We now use the lens and source redshifts and number densities described in the previous subsection to make simulated stacked shear and flexion maps with uncertainties. We fit unknown lens parameters to these maps to obtain probability distributions as a function of lens ellipticity. We compare the results from shear, one-flexion and three-flexion. We also investigate how our results depend on the survey depth.

The final input to our analysis is an estimate of the uncertainty on the shear and flexion measurements for a single background galaxy. We assume the uncertainty on the shear measurement for each galaxy is 0.3 for each shear component. Although this number does depend on the depth of the survey, for deep surveys it is dominated by the intrinsic distribution of galaxy ellipticities which contribute about 0.2 to the shears.
Owing to the extreme dificulty in obtaining flexion measurements from real data \citep{Rowe}
we have used a conservative estimate of $\sigma_{\mathcal{F}}=\sigma_{\mathcal{G}}=0.1 $. This is larger than the estimated noise on the flexions $\sigma_{\mathcal{F}} = 0.03\,\, $ and $\sigma_{\mathcal{G}} = 0.04\,\, $ used in \citet{GoldbergLeonard}. Again it is known that noise on the flexion is dependent on survey depth. 
These uncertainties are divided by the square root of the number of galaxies in each pixel on the collated postage stamp to obtain an estimate of the noise
for that pixel.

We consider trial values of the lens mass and lens ellipticity and compute predicted shear and flexion maps. We keep the lens centered on the origin and the orientation along the $x$-axis since this would be implicit in the stacking procedure.
We compute the probability of each model by calculating a $\chi^2$ between the trial map and the fiducial map, using the uncertainties computed for each pixel. We marginalise over the mass to find the probability as a function of lens ellipticity and find the uncertainty on the ellipticity from the 68 per cent confidence limits. This is illustrated in Fig.~\ref{likeEUCLID} for the fiducial space-based survey.
The constraints on ellipticity from the space based survey are $\sigma_e =2.0 \times10^{-5}$ from one-flexion, $\sigma_e=9.1\times10^{-6}$ from three-flexion and $\sigma_e =4.5 \times10^{-3}$ from shear.
It is clear that flexion is a much more powerful method with which to constrain lens ellipticity than shear alone. Indeed it is not even convenient to show the constraints from shear on the same axes.

\begin{figure}
\psfrag{mathcalF}{$\mathcal{F}$}
\psfrag{mathcalG}{$\mathcal{G}$}
\psfrag{sigmae}{$\sigma_e$}
\psfrag{Magnitude Limit}{$r_{lim}$}
\includegraphics[width=8cm]{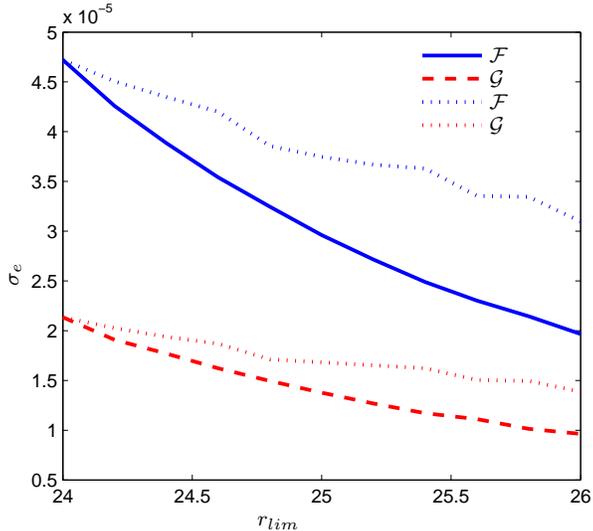}
\caption{Error on ellipticity with respect to magnitude limit of a ground based survey.
The dotted lines show constraints obtained using a varying lens population whilst the blue solid line and red dashed line show constraints using the lens population from the $r_{lim}=24$ ground based survey. As the magnitude limit of the survey increases so the uncertainty in the mean ellipticity of galaxy dark matter haloes decreases.
There is about a factor of two improvement in moving from an $r_{lim}=24$ to an $r_{lim}=26$ survey.}
\label{maglim}
\end{figure}

We show the uncertainties in the lens ellipticity using flexion estimated for ground-based surveys as a function of survey depth in Fig.~\ref{maglim}.
We see that the two types of flexion give similar size constraints, with the three flexion giving slightly tighter constraints.
When the lens population is kept fixed as the survey depth is increased (solid lines) the uncertainty approximately halves as the magnitude is increased by two. When the lens population is varied to keep the number density of lenses constant the constraints vary little. This is because, as discussed earlier, the lenses become closer and lower mass.
>From our $r_{lim}=24$ survey $\sigma_e = 4.7 \times10^{-5}$ with one-flexion and $\sigma_e = 4.4\times10^{-5}$ with three flexion. This compares to $\sigma_e = 8.2 \times 10^{-3}$ using the shear. From our $r_{lim}=26$ survey constraints are $\sigma_e = 2.0\times10^{-5}$ with one flexion and $\sigma_e = 9.7\times10^{-6}$ with three flexion. Shear provides constraints of $\sigma_e =4.4\times10^{-3}$.
These numbers are very similar to those from the space-based survey.

\begin{figure}\label{erver}
\psfrag{sigmatag}{$\sigma_{\mathcal{F}}/\sigma_{\mathcal{G}}/arcsec^{-1}$}
\psfrag{mathcalF}{$\mathcal{F}$}
\psfrag{mathcalG}{$\mathcal{G}$}
\psfrag{sigmae}{$\sigma_e$}
\includegraphics[width=8cm]{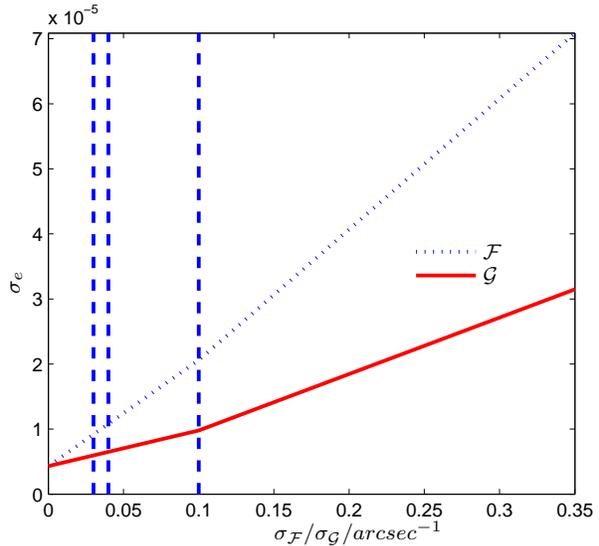}
\caption{Error on dark matter halo ellipticity as a function of error on flexion measurements for one-flexion (dotted line) and three-flexion (solid line). This is for our space based survey. The estimated error on the flexions of $\sigma_{\mathcal{F}}=0.03$ and $\sigma_{\mathcal{G}}=0.04$ consistent with \citet{Bacon}, \citet{GoldbergLeonard} and \citet{Rowe} are illustrated, as is the value we have used of $\sigma_{\mathcal{F}}=\sigma_{\mathcal{G}}=0.1$.
}
\end{figure}

In Fig.~\ref{erver} we illustrate the ellipticity uncertainty as a function of assumed flexion uncertainty.
This shows that the relative error on the flexion measurements themselves can be large without compromising flexion's ability to constrain the ellipticity of dark matter haloes. It has been observed that at fainter magnitudes the uncertainty in flexion measurements is greater, this effect is most noticeable for three-flexion (David Bacon private communication). If, for example, the uncertainty in one-flexion measurements were $\sim0.1$ and in three-flexion $\sim0.3$ then the constraints obtained from each would be roughly the same.

\section{CONCLUSION} \label{conclusion}

We have generalised the equations for shear from an ellipsoidal
mass distribution to flexion. These could in principle be extended
to even higher order lensing effects by further differentiation.
This differs from previous work by~\citet{Baltz} who approximate an ellipsoidal
mass distribution with a sum of ellipsoidal potentials.

We have demonstrated that galaxy-galaxy flexion can constrain the ellipticity of galaxy dark matter haloes up to two orders of magnitude better than shear. Such precision measurements would be very useful in constraining models of galaxy formation and galaxy dark matter. Increasing the magnitude limit of a ground based survey from $r_{lim}=24$ to $r_{lim}=26$ tightens the constraints on ellipticity by up to a factor of two.

The exact values depend heavily on the assumed intrinsic flexion uncertainty coming from the intrinsic galaxy shapes. As mentioned in section \ref{constraints on lens ellipticity} the flexion uncertainty is expected to increase with survey depth, this could be accounted for with a more detailed treatment. This will make our predicitions for ellipticity constraints slightly optimistic. Furthermore, we have not considered the CCD resolution for the space based survey. Pixels much larger than an HST style camera could substantially increase flexion uncertainty. Additionally, we have assumed Gaussian errors on the flexion measurements yet it is known that observed flexion errors are far from Gaussian~\citep{Rowe}.

In carrying out a galaxy halo ellipticity observation there is
considerable freedom in selecting the lens and source population
to include in the analysis. Since we are interested in a raw
comparison between shear and flexion we used a simple redshift
cut on the lens population. However it may be useful to split the lens
population by observed ellipticity before stacking~\citep{Hoekstra:2003pn,Mandelbaum,Parker},
to increase the signal-to-noise of any detection of a difference
between the halo and light ellipticities.
Furthermore it is of interest to study the evolution of halo
ellipticity with halo mass and redshift~\citep{kasun05, Allgood2006,hopkins05,ho06}.

In practice the selection of lens and/or source galaxies
may be carried out using photometric redshifts, which would not
allow a straight cut in redshift, but would cause the two populations
to overlap somewhat. This must be carefully accounted for in determinations
of the mass, but has a second order effect on the ellipticity.

We have further assumed that galaxies at all positions behind
the lenses may be used for shear and flexion analysis. However
it will be impossible to measure background galaxy shapes very
close to the foreground galaxy, due contamination of the
measurement by light from the foreground object. Depending on
the ellipticity and orientation of the light relative to the
mass, this will impose a roughly elliptical mask on the
background galaxy catalogue which will weaken both the
constraints on the halo ellipticity for both shear and flexion
measurements. The exact extent will depend on the resolution of
the telescope and the efficiency of the shear or flexion
measurement code to deal with varying background levels across
the image of a lensed galaxy, which is not yet widely
established.

Current surveys such as the Hubble Space Telescope (HST) Cosmic Evolution Survey (COSMOS) and the Canada-France-Hawaii Telescope Legacy Survey (CFHTLS) might already be able to place tight constraints on halo ellipticity using flexion, if the flexions can be measured with sufficient accuracy.
We look forward to future measurements from high precision imaging surveys such as the Dark Energy Survey (DES), Pan-STARRS, the Large Synoptic Survey Telescope (LSST) and a space mission such as Euclid and/or the Joint Dark Energy Mission (JDEM).

\section*{Acknowledgements}

Many thanks to Lisa Voigt and David Bacon for useful discussions, to Barney Rowe for lending his thesis, to Alex Refregier and all at CEA Saclay where part of this work was carried out.
SLB acknowledges support from a Royal Society University Research Fellowship.

\bsp

\bibliographystyle{mn2e}

\label{lastpage}

\end{document}